\documentclass{optica-article}

\journal{opticajournal} 

\articletype{Research Article}

\usepackage{xcolor}
\usepackage{cleveref}
\usepackage{amsmath}

\usepackage{amssymb}
\usepackage{upgreek}
\usepackage{standalone}

\newcommand{\red}[1]{\textcolor{red}{#1}}
\newcommand{\um}{$\upmu$m}
\newcommand{\ftwof}{$f$-$2f$}
\newcommand{\bint}{$\beta_\mathrm{int}$}

\crefname{figure}{Fig.}{Figs.}
\Crefname{figure}{Figure}{Figures}

\begin{document}

\title{Picojoule-level self-referencing with dual dispersive waves in thin-film lithium niobate}

\author {
Kiyoung Ko, \authormark{1} 
Ian Christen, \authormark{1} 
Kamila Kunes, \authormark{1}
Clayton Cheung, \authormark{1}
Chun-Ho Lee, \authormark{1, 2}
Kiwon Kwon, \authormark{1}
Mihir Chaudhari, \authormark{1}
Ran Yin, \authormark{1}
Tushar Sanjay Karnik \authormark{1, 2}
and Mengjie Yu \authormark{1,3*}
}

\address{\authormark{1}Department of Electrical Engineering and Computer Sciences, University of California, Berkeley, Berkeley, CA, 94720, USA\\
\authormark{2}Opticore Inc., Berkeley, CA, 94704, USA\\
\authormark{3}Material Sciences Division, Lawrence Berkeley National Laboratory, Berkeley, CA, 94720, USA}

\email{\authormark{*} mengjie.yu@berkeley.edu;} 


\begin{abstract*}
We demonstrate pulse-energy-efficient \ftwof\ self-referencing of telecom-band pulses in a monolithic thin-film lithium niobate (TFLN) waveguide that combines Kerr spectral broadening with quasi-phase-matched (QPM) second-harmonic generation (SHG). Dispersion engineering generates dual dispersive waves (DWs) near 1 and 2 \um, placed symmetrically around the pump to reduce the required spectral broadening. With an on-chip pulse energy of only 16.8 pJ, we detect a carrier-envelope-offset frequency ($f_\text{ceo}$) beatnote with 32 dB signal-to-noise ratio (SNR), and at 30 pJ we stabilize the $f_\text{ceo}$ of an Er-fiber laser which is, to our knowledge, the lowest pulse energy reported for $f_\text{ceo}$ stabilization at telecom wavelengths. Compared with an identical waveguide without periodic poling, the QPM section enhances the beatnote SNR by 43 dB and reduces the required pulse energy fourfold. We further show that a mode-crossing dispersive wave (DW), a chirped poling period, and mitigating temporal walk-off by aligning the poling with the onset of DW generation are key to this performance.
\end{abstract*}

\section{Introduction}
Self-referencing with an \ftwof\ interferometer is key to detecting and stabilizing the carrier-envelope-offset frequency ($f_\text{ceo}$) of a pulse train, and is thus critical for fully stabilized optical frequency combs used in frequency metrology, optical atomic clocks, precision spectroscopy, and attosecond science \cite{udem_optical_2002, diddams_evolving_2010, cundiff_colloquium_2003, cormier_self-referencing_2005}. Energy-efficient $f_\text{ceo}$ stabilization is becoming increasingly important with the miniaturization of pulse-generation hardware. For instance, mode-locked erbium (Er)-doped fiber lasers provide a mature platform for compact frequency combs in the telecommunication band near 1.5 \um, and self-referencing directly from their oscillator output would eliminate a dedicated optical amplifier, significantly reducing system size, electrical power consumption, and heat load \cite{wu_amplifier-free_2026, manurkar_fully_2018}. This imposes a tight energy budget: a typical oscillator power of 10 mW at 100 MHz corresponds to only 100 pJ per pulse. The constraint is even more severe for high-repetition-rate sources, since at a fixed average power $P_\text{avg}$ the pulse energy $E_p = P_\text{avg}/f_\text{rep}$ scales inversely with the repetition rate $f_\text{rep}$. Integrated pulse and comb sources naturally operate at high repetition rates owing to their compact device size \cite{kippenberg_dissipative_2018}; recent examples generate ultrashort pulses with energies of only a few picojoules at 10--30 GHz \cite{ren_few-cycle_2026, yu_integrated_2022, guo_ultrafast_2023}. Reaching 100 pJ per pulse at these rates would require 1--3 W of average power, which challenges the power handling of integrated waveguides. Reducing the pulse energy required for self-referencing is therefore essential for both compact fiber-based and integrated frequency comb sources.

Conventional \ftwof\ interferometers cascade supercontinuum generation (SCG) in a Kerr ($\chi^{(3)}$) medium with second-harmonic generation (SHG) in a separate bulk crystal, such as periodically poled lithium niobate (PPLN) \cite{klenner_gigahertz_2016, lamb_optical-frequency_2018} or potassium titanyl phosphate (PPKTP) \cite{carlson_self-referenced_2017}, often with a delay stage to temporally overlap the $f$ and $2f$ components; the resulting interface losses further strain the energy budget. Benefiting from nonlinear interactions enhanced by tight optical confinement, integrated photonics has enabled chip-based $f_\text{ceo}$ detection on silicon nitride (SiN) \cite{carlson_self-referenced_2017, klenner_gigahertz_2016, okawachi_carrier_2018}, aluminum nitride (AlN) \cite{hickstein_ultrabroadband_2017}, gallium nitride (GaN) \cite{volpini_cladding_2026}, thin-film lithium niobate (TFLN) \cite{yu_coherent_2019, okawachi_chip-based_2020, chi_chip-based_nodate, tang_-chip_2026, wu_visible--ultraviolet_2024}, and heterogeneous $\mathrm{Ta_2O_5}$/TFLN \cite{wu_amplifier-free_2026,brodnik_monolithic_2026}, but achieving efficient spectral broadening, SHG, and temporal overlap simultaneously at low pulse energy remains difficult. Existing Kerr-broadening designs typically rely on a dispersive wave (DW) near 780 nm, a full octave above the pump frequency \cite{volpini_cladding_2026, chi_chip-based_nodate, tang_-chip_2026, brodnik_monolithic_2026}. SHG is often limited by a weak $\chi^{(2)}$ nonlinearity or non-ideal phase-matching conditions \cite{okawachi_carrier_2018, hickstein_ultrabroadband_2017, volpini_cladding_2026, chi_chip-based_nodate, carlson_self-referenced_2017}. Recent work has instead used group-velocity-matched (GVM) $\chi^{(2)}$ broadening in PPLN waveguides to lower the detection energy to about 10 pJ, without stabilization, but the GVM condition is achieved by operating with a central wavelength near 2 \um\ which comes with its own challenges \cite{jankowski_ultrabroadband_2020, jankowski_supercontinuum_2023, hamrouni_picojoule-level_2024}; chirped PPLN has enabled detection via higher-order harmonics at about 90 pJ \cite{wu_visible--ultraviolet_2024}. To date, chip-based $f_\text{ceo}$ stabilization at 1.5 \um\ has been demonstrated only with pulse energies above 100 pJ, in SiN \cite{carlson_self-referenced_2017}, TFLN \cite{okawachi_chip-based_2020}, and AlN \cite{hickstein_ultrabroadband_2017}. Therefore, a low-energy approach directly compatible with compact telecom-band pulse sources is still missing.

Here, we demonstrate energy-efficient $f_\text{ceo}$ detection and stabilization near 1.5 \um\ by co-designing spectral broadening, quasi-phase-matched (QPM) SHG, and temporal overlap within a single TFLN waveguide. TFLN combines strong $\chi^{(2)}$ and $\chi^{(3)}$ nonlinearities with ferroelectric domain engineering for efficient SHG while leaving the waveguide geometry free for dispersion engineering. We use this freedom to generate octave-separated DWs near 1 and 2 \um, positioned symmetrically around the pump wavelength, significantly reducing the Kerr broadening required to reach the $f$ and $2f$ components compared with previous designs \cite{volpini_cladding_2026, chi_chip-based_nodate, tang_-chip_2026, brodnik_monolithic_2026}. The 18-mm-long device (21-mm including tapered input and output) comprises a 13-mm unpoled section for Kerr broadening followed by a 5-mm periodically poled section that frequency-doubles the long-wavelength DW. Positioning the poled section near the onset of DW generation mitigates temporal walk-off between the octave-separated DWs, providing sufficient temporal overlap without an external delay line. With an on-chip pulse energy of only 16.8 pJ, we detect an $f_\text{ceo}$ beatnote with a 32 dB signal-to-noise ratio (SNR) and use it to stabilize the $f_\text{ceo}$ of the input pulse train. Moreover, compared with a reference waveguide without a PPLN section, we demonstrate a 43 dB improvement in the $f_\text{ceo}$ beatnote SNR and a fourfold reduction in the required pulse energy. We further investigate how avoided mode crossings and the PPLN design, including its poling period, chirp, and position, affect the $f_\text{ceo}$ beatnote. To our knowledge, this work is the first demonstration of $f_\text{ceo}$ detection and stabilization using this scheme, with the lowest pulse energy reported for $f_\text{ceo}$ detection near 1.5 \um, and a more than sixfold reduction relative to previous demonstrations of $f_\text{ceo}$ stabilization in this wavelength range \cite{carlson_self-referenced_2017, okawachi_chip-based_2020, hickstein_ultrabroadband_2017}.

\section{Device design and characterization}

\begin{figure}[htbp]
\centering\includegraphics[width=9cm]{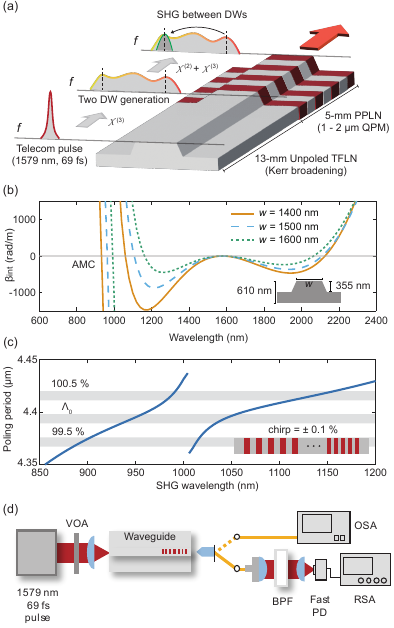}
\caption{
Concept and design of the proposed self-referencing waveguide.
(a) Schematic of the two-section waveguide, which generates two DWs and frequency-doubles the long-wavelength DW to obtain the $f_\text{ceo}$ beatnote. (b) Integrated dispersion (\bint) of TFLN waveguides with varying top widths ($w$) simulated using a Tidy3D mode solver. A total film thickness of 610 nm, an etch depth of 355 nm, and a sidewall angle of $59^{\circ}$ were used to match the fabricated devices. The discontinuities near 1 \um\ are due to the avoided mode crossing (AMC). (c) SHG phase-matching curve calculated for $w$ = 1524 nm, the width used in all subsequent experiments unless otherwise noted. The gray shaded regions indicate three distinct periodic poling configurations applied to the waveguide, with center periods of 99.5, 100, and 100.5 \% of $\Lambda_0$ = 4.3936 \um\ and a chirp of $\pm0.1$ \%. The AMC causes the curve to diverge near 1 \um. (d) Schematic drawing of the measurement setup. VOA: variable optical attenuator, BPF: band-pass filter, OSA: optical spectrum analyzer, PD: photodetector, RSA: RF spectrum analyzer.
}
\label{fig:setup n waveguide design}
\end{figure}

The design of the $x$-cut MgO-doped TFLN waveguide was optimized to perform the sequential third- and second-order nonlinear processes shown in \cref{fig:setup n waveguide design}a. The 13-mm-long first section, responsible for Kerr-induced spectral broadening, was intentionally left unpoled to avoid depleting the pump through SHG, while the 5-mm-long second section was periodically poled with a period near 4.4 \um\ to frequency-double the DW near 2 \um\ to the 1 \um\ region.
The waveguide width ($w$) was chosen to produce two near-zero crossings of the integrated dispersion (\bint) at the desired DW wavelengths, simulated using MIT Photonic-Bands (MPB) and Tidy3D mode solver based on a reference material index data \cite{gayer_temperature_2008}. \cref{fig:setup n waveguide design}b shows \bint\ for three top widths, calculated for the fabricated cross section with a film thickness of 610 nm and an etch depth of 355 nm (inset of \cref{fig:setup n waveguide design}b). Waveguides with a range of widths were fabricated to account for fabrication variations. The poling period $\Lambda_0$ = 4.3936 \um\ required for SHG was calculated from the same simulation, and nine poling configurations, combining three center periods (99.5, 100, and 100.5 \% of $\Lambda_0$) and three chirps (0, $\pm$0.1, and $\pm$0.2 \%), were designed to test the effect of the poling parameters on the $f_\text{ceo}$ beatnote.

The TFLN film was first periodically poled using a custom automated probing setup, after which the ridge waveguides were defined by electron-beam lithography using ma-N resist and inductively coupled plasma dry etching. Reference waveguides with identical geometry but without a poled section were fabricated adjacent to each poled waveguide. The supercontinuum spectra of the reference waveguides agree well with simulations (Supplement 1, Section 1), which predict the desired DW positions for a top width of 1524 nm; waveguides with this width were used in all subsequent experiments for $f_\text{ceo}$ detection and stabilization. The corresponding second-harmonic (SH) phase-matching curve is shown in \cref{fig:setup n waveguide design}c.

A distinctive feature arising from the avoided mode crossing (AMC) between the $\mathrm{TE_{00}}$ and $\mathrm{TM_{10}}$ modes of the birefringent $x$-cut TFLN film \cite{hamrouni_picojoule-level_2024} is visible in both the \bint\ and SH phase-matching curves (\cref{fig:setup n waveguide design}b and c). In the dispersion, the AMC creates a third zero crossing of \bint\ slightly below 1 \um. Consequently, a strong dispersive wave, hereafter referred to as the mode-crossing dispersive wave (MCDW), is expected in addition to the short- and long-wavelength dispersive waves (SWDW, LWDW) at 1.1 and 2.2 \um. The AMC also causes a divergence in the SH phase-matching curve at the SHG wavelength of 1 \um\ (\cref{fig:setup n waveguide design}c), splitting the QPM band of the PPLN into two separate spectral regions. These AMC-induced features strongly affect the spectral dependence of the $f_\text{ceo}$ beatnote intensity, as observed experimentally in Section 3.

The fabricated waveguides were characterized as follows using a setup presented in \cref{fig:setup n waveguide design}d. An Er-doped fiber laser (C-Fiber, Menlo Systems) with a center wavelength of 1579 nm, a repetition rate of 100 MHz, and a pulse duration of 69 fs was used as the pump. The pump pulses were coupled into the waveguide in TE polarization using a free-space objective, and the output was collected with a lensed fiber. For spectral measurements, the output was sent to an optical spectrum analyzer (OSA); for $f_\text{ceo}$ measurements, it was passed through a band-pass filter (BPF) to a photodetector (PD; FPD510-FS-NIR, Menlo Systems) connected to an RF spectrum analyzer (RSA). The BPF, composed of an adjustable slit and a grating, selects the spectral components that contribute to the $f_\text{ceo}$ beatnote.

\section{Near-two-octave-spanning spectrum and spectral dependence of the $f_\text{ceo}$ beatnote}

\begin{figure}[htbp]
\centering\includegraphics[width=\linewidth]{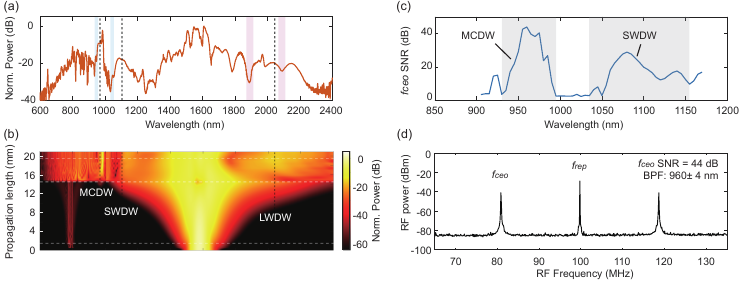}
\caption{
Near-two-octave-spanning spectrum and spectral dependence of the $f_\text{ceo}$ beatnote.
(a) Measured supercontinuum spectrum at $E_p \approx 23.6$ pJ for the waveguide with $w$ = 1524 nm, a poling period of $\Lambda_0$, and a $\pm0.1$ \% chirp. The vertical dashed lines indicate the three DW wavelengths. The QPM bandwidth and the corresponding first-harmonic wavelengths are marked with blue and pink shaded regions. (b) Spectral evolution simulated using the generalized nonlinear envelope equation (GNEE). Horizontal dashed lines indicate the boundaries between waveguide sections: 1.5-mm tapered input ($w$ = 3 \um), 13-mm unpoled section ($w$ = 1524 nm), 5-mm PPLN section ($w$ = 1524 nm), and 1.5-mm tapered output identical to the input. The figure shares its x-axis with (a). (c) $f_\text{ceo}$ beatnote SNR versus the center wavelength of the $\sim$8 nm BPF applied to the spectrum in (a). (d) RF spectrum at the SNR maximum in (c), showing the $f_\text{ceo}$ beatnote near 80 MHz with 44 dB SNR and the $f_\text{rep}$ signal at 100 MHz with 56 dB SNR.
} 
\label{fig:fceo SNR spectrum}
\end{figure}

Among the fabricated devices, the waveguide with a poling period of $\Lambda_0$ and a $\pm0.1$ \% chirp showed the most efficient $f_\text{ceo}$ beatnote generation. \cref{fig:fceo SNR spectrum}a shows its output supercontinuum spectrum measured at an on-chip pulse energy of $E_p$ = 23.6 pJ, which agrees well with the output of the spectral evolution simulated at the same pulse energy using the generalized nonlinear envelope equation (GNEE) \cite{genty_nonlinear_2007, wabnitz_harmonic_2010} (\cref{fig:fceo SNR spectrum}b). As shown in \cref{fig:fceo SNR spectrum}b, the pulse spectrum broadens through the Kerr nonlinearity as it propagates along the 13-mm unpoled section. A weak SH signal near 790 nm is also generated by the intrinsic, non-phase-matched SHG of the pump, but it is too weak to affect the Kerr broadening. Near the end of the unpoled section, the three DWs begin to form as the pump coherently transfers energy to their phase-matched wavelengths, and they continue to grow in the PPLN section as the spectrum broadens further to span nearly two octaves, from 600 to 2300 nm. The three DW wavelengths (MCDW, SWDW, and LWDW) and the QPM band are marked as dashed lines and shaded regions, respectively, for comparison with the spectra.

To confirm that the $f_\text{ceo}$ beatnote arises from interference between the DWs near 1 \um\ and the SH of the DW near 2 \um, we measured the spectral dependence of its SNR by passing the spectrum in \cref{fig:fceo SNR spectrum}a through the tunable BPF and monitoring the $f_\text{ceo}$ signal on the RSA. The BPF has a bandwidth of $\sim$8 nm, and the RSA resolution bandwidth (RBW) and video bandwidth were both set to 10 kHz. As shown in \cref{fig:fceo SNR spectrum}c, the beatnote is generated by wavelength components between 900 and 1150 nm, and its SNR peaks precisely at the MCDW (960 nm) and SWDW (1080 nm) wavelengths. In contrast, the SNR is minimized between 995 and 1035 nm, consistent with the spectral gaps in both DW generation and SHG predicted from the AMC-induced phase mismatch. 

Moreover, the beatnote SNR is higher near the MCDW than near the SWDW (\cref{fig:fceo SNR spectrum}c). We attribute this mainly to two factors: the higher spectral power at the fundamental wavelength responsible for SHG at the MCDW compared to that at the SWDW before entering the PPLN, and the greater spectral overlap of the MCDW with the generated SH. These effects effectively compensate for the reduced nonlinear efficiency induced by mode hybridization \cite{hamrouni_picojoule-level_2024}, generating a stronger signal in both the optical and $f_\mathrm{ceo}$ SNR spectra. 
This explanation is further supported by the simulated SCG spectra before and after the PPLN presented in Supplement 1, Section 2, which demonstrate that although the spectral power near MCDW is weaker than near the SWDW prior to the PPLN section, it becomes stronger after the PPLN. Notably, this trend persisted regardless of the poling period or chirp, even in a polling period of 100.5\% $\times$ $\Lambda_0$ where the spectral overlap near the SWDW is greater than near the MCDW with a lesser difference between their SNR. This indicates that the spectral power at the SHG fundamental wavelength currently serves as the dominant factor determining the $f_\mathrm{ceo}$ beatnote SNR. 

\cref{fig:fceo SNR spectrum}d shows the RF spectrum with the BPF centered at 960 nm, where the $f_\text{ceo}$ beatnote is maximized, showing an $f_\text{ceo}$ beatnote near 80 MHz with 44 dB SNR. The $f_\text{ceo}$ beatnote is only 12 dB lower than the $f_\text{rep}$ signal at 100 MHz, indicating that the filtered spectral components contribute efficiently to the $f_\text{ceo}$ beatnote.

\section{Pulse-energy-efficient $f_\text{ceo}$ detection}

\begin{figure}[htbp]
\centering\includegraphics[width=0.7\linewidth]{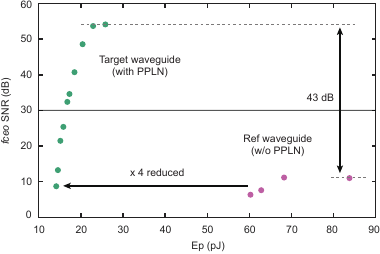}
\caption{
Power dynamics of $f_\text{ceo}$ beatnote SNR measured on the optimized waveguide with a PPLN and a reference waveguide without it. Spectrum between 950-1000 nm was band-pass filtered to maximize the SNR. The periodically poled waveguide requires four times less pulse energy for the onset of the $f_\mathrm{ceo}$ beatnote and shows 43 dB stronger SNR. 
}
\label{fig:power dynamics poled n unpoled compared}
\end{figure}

\begin{figure}[htbp]
\centering\includegraphics[width=\linewidth]{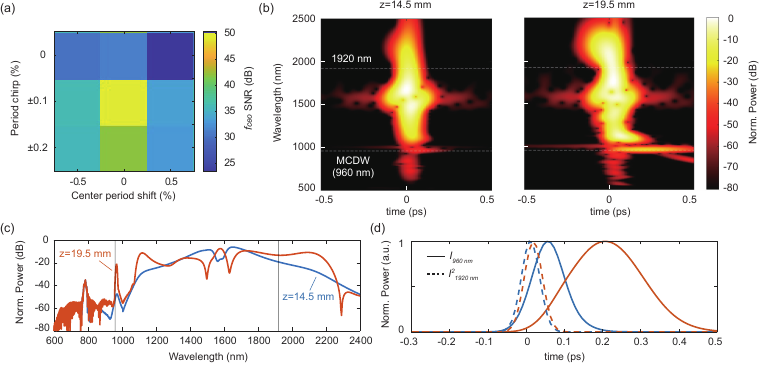}
\caption{
(a) Maximum $f_\text{ceo}$ SNR measured on nine waveguides with nearly identical cross sections but different center poling period and chirps. (b) Supercontinuum spectrogram simulated on the reference waveguide without a PPLN. The propagation direction coordinate ($z$) of 14.5 and 19.5 mm corresponds to where the 5 mm PPLN starts and ends on the periodically poled waveguides. Temporal Gaussian envelope with FWHM of 50 fs was used as a filtering function of the spectrogram.
(c) Supercontinuum spectrum corresponding to the two z positions shown in (b). (d) Normalized temporal intensity envelope of the MCDW ($I_{960\: \mathrm{nm}}$, solid lines) and the corresponding first-harmonic components (1920 nm, dashed lines) at the two z positions showing temporal walk-off effect. The envelopes were obtained by filtering 930 to 1000nm and 1850 to 2000 nm spectral components from the spectrogram in (b). The displayed trace for 1920 nm is the square of its temporal intensity envelope ($I^2_{1920 \:\mathrm{nm}}$), which approximates the intensity of second harmonic generated by it.
}
\label{fig:spectra power dynamics simulation}
\end{figure}

Next, we show that monolithically integrating the PPLN directly after the spectral broadening section greatly enhances the $f_\text{ceo}$ beatnote and lowers the required pulse energy. The impact of the PPLN segment is clearly shown in \cref{fig:power dynamics poled n unpoled compared}, where the $f_\text{ceo}$ beatnote SNR of the waveguide with PPLN section is plotted against the pulse energy and compared with that of an adjacent reference waveguide without the PPLN section. For both devices, the SCG spectrum was filtered to 950--1000 nm, which maximized the detected SNR. With the PPLN, the $f_\text{ceo}$ beatnote emerges above the noise floor at $E_p$ = 14 pJ, and its SNR rises rapidly from 8.7 dB to 54 dB as $E_p$ increases from 14 to 26 pJ. As a reference point, we use an SNR of 32 dB at $E_p$ = 16.8 pJ, which is sufficient for a stable $f_\text{ceo}$ lock, as demonstrated below, and is used for comparison with previous work in \cref{tab:device comparison v2}. The reference waveguide also produces an $f_\text{ceo}$ beatnote, but only at much higher pulse energies: the beatnote appears at 60 pJ, and its SNR saturates at about 11 dB beyond 70 pJ. This is due to the non-ideal phase matching of SHG without poling, which also limited the pulse energy required in previous non-phase-matched self-referencing demonstrations \cite{okawachi_chip-based_2020}. Overall, the PPLN enhances the maximum SNR by 43 dB and reduces the pulse energy required for $f_\text{ceo}$ detection by a factor of four. The 16.8 pJ pulse energy corresponds to an average power of only 1.68 mW at the 100 MHz repetition rate of the Er-fiber laser used here, well within the output of typical Er-fiber oscillators, and scales to the 150--500 mW range at 10--30 GHz, offering a path toward self-referencing of integrated high-repetition-rate pulse sources on the same TFLN platform.

\section{Effect of PPLN parameters on beatnote generation}

We further investigated the effect of the poling parameters by comparing the $f_\text{ceo}$ SNR of the nine waveguides with identical target geometries but different poling configurations, i.e., different center periods and chirps (\cref{fig:spectra power dynamics simulation}a). All waveguides were pumped at the same off-chip power of 9.7 dBm, corresponding to $E_p \approx 21$ pJ on chip. For every waveguide, the highest SNR was obtained with the BPF set within 950--1000 nm, where the MCDW is located. The achievable SNR depends on both the center period and the chirp, varying by more than 25 dB owing to changes in the spectral overlap between the DWs and the SH signal. Unchirped waveguides showed the lowest SNR, indicating that chirping is beneficial for efficient frequency doubling of the broadband DW. This sensitivity to the poling period also suggests that adaptive poling and precision alignment of the SHG wavelength could further enhance the SNR. 

The position and length of the PPLN section also affect $f_\text{ceo}$ generation because of the temporal walk-off between the $f$ and $2f$ spectral components. \cref{fig:spectra power dynamics simulation}b shows simulated spectrograms of the supercontinuum in the unpoled reference waveguide at $z$ = 14.5 and 19.5 mm, corresponding to the start and end of the 5-mm PPLN section in the poled waveguides. As seen in the spectrograms and the corresponding spectra (\cref{fig:spectra power dynamics simulation}c), the MCDW (960 nm) begins to emerge around $z$ = 14.5 mm. This is the optimal position to start the PPLN section, because frequency doubling before the DWs form cannot contribute to the beatnote. As the pulse propagates further, the MCDW grows by 27 dB and saturates by $z$ = 19.5 mm. Meanwhile, the 930--1000 nm and 1860--2000 nm components walk off from each other (\cref{fig:spectra power dynamics simulation}d): for a group-velocity mismatch of 75 fs/mm, the relative delay accumulated over the 5-mm PPLN ($\sim$375 fs) well exceeds the $\sim$150 fs duration (FWHM) of the MCDW. Extending the PPLN beyond 5 mm would therefore not significantly strengthen the $f_\text{ceo}$ signal. A more rigorous analysis of the walk-off effect is given in Supplement 1, Section 3. These results establish a simple design rule for monolithic \ftwof\ waveguides: the PPLN section should begin where the MCDW emerges and end where the DW growth saturates and the $f$ and $2f$ components walk off. Our device operates close to this optimum, which contributes directly to $f_\text{ceo}$ detection at only 16.8 pJ.

\section{$f_\text{ceo}$ stabilization via self-referencing}

\begin{figure}[htbp]
\centering\includegraphics[width=8.5cm]{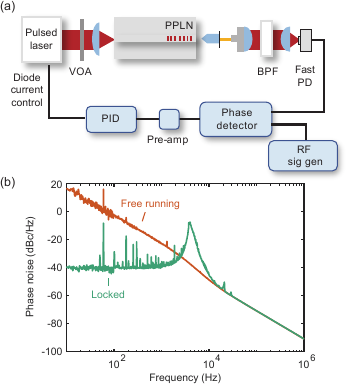}
\caption{
$f_\text{ceo}$ stabilization via self-referencing.
(a) Schematic of the setup used for locking the $f_\text{ceo}$ of the input pulsed laser. RF sig gen: reference RF signal generator; PID: proportional--integral--derivative controller. (b) Measured phase-noise spectra of the $f_\text{ceo}$ beatnote in the free-running and locked cases. When locked, the phase noise below 30 kHz is suppressed to the $-40$ dBc/Hz level.
}
\label{fig:fceo locking}
\end{figure}

Finally, we demonstrate $f_\text{ceo}$ stabilization using the TFLN device, with the locking setup shown in \cref{fig:fceo locking}a. The $f_\text{ceo}$ beatnote and a reference RF signal are sent to a phase detector (DXD212, Menlo Systems), which converts their phase difference into an error signal. This error signal drives a proportional--integral--derivative (PID) controller (LaserLock, TEM Messtechnik, 100 kHz bandwidth) that feeds back to the pump-diode current of the laser. For locking, we used a slightly higher pulse energy of $E_p \approx 30$ pJ to reach the $-30$ dBm RF power required by the phase detector; this requirement could be relaxed by adding an RF amplifier. As shown in \cref{fig:fceo locking}b, locking suppresses the phase noise below 30 kHz to the $-40$ dBc/Hz level relative to the free-running case. The peak near 30 kHz is the servo bump of the feedback loop, which marks its locking bandwidth. This bandwidth is limited by the slow response of the pump-current actuator, which is governed by the gain dynamics of the Er-doped fiber, rather than by the TFLN device.

\section{Discussion and Conclusion}

We demonstrated $f_\text{ceo}$ detection and stabilization of telecom-band pulses from an Er-fiber laser using a monolithic periodically poled TFLN waveguide. To our knowledge, this is the first $f_\text{ceo}$ generation scheme that combines dual DWs near 1 and 2 \um\ with QPM SHG, and it achieves the lowest pulse energy reported for $f_\text{ceo}$ stabilization at telecom wavelengths: an $f_\text{ceo}$ beatnote with 32 dB SNR at only 16.8 pJ, and a stabilized $f_\text{ceo}$ at 30 pJ with the phase noise below 30 kHz suppressed to the $-40$ dBc/Hz level. Compared with an unpoled reference waveguide, the PPLN section enhanced the beatnote SNR by 43 dB and reduced the onset pulse energy by a factor of four. More broadly, this work establishes a device architecture that simultaneously engineers octave-spanning Kerr broadening, QPM SHG, and the temporal dynamics of the $f$ and $2f$ components: the AMC-induced MCDW, a chirped poling period, and a PPLN section positioned at the onset of DW generation together minimize the required pulse energy. \Cref{tab:device comparison v2} compares this work with previous monolithic waveguides for $f_\text{ceo}$ detection.

The TFLN platform is particularly attractive in this regard: the pulse energy demonstrated here is approaching the range produced by recent on-chip pulse generators on the same platform \cite{ren_few-cycle_2026, yu_integrated_2022, guo_ultrafast_2023}, which could enable self-referenced pulsed-light systems without heterogeneous integration of separate nonlinear stages. Further reductions in pulse energy and waveguide length could benefit from recent advances in supercontinuum waveguide design, such as width-modulated waveguides \cite{carlson_generating_2019, arduin_doubling_2026} and inverse design \cite{lee_inverse_2024}, paving the way toward fully integrated, self-referenced pulsed-light systems on a single chip.

\begin{table}
    \centering
    \small
    \setlength{\tabcolsep}{3pt}
    \begin{tabularx}{\linewidth}{cc>{\centering\arraybackslash}Xccccc}
    \hline\hline
        Ref. & Platform & SHG mechanism & $\lambda_p$ (nm) & $\tau$ (fs) & $E_p$ (pJ) & SNR (dB) & $f_\text{ceo}$ lock \\ \hline
        
        \cite{okawachi_carrier_2018} & $\mathrm{Si_3N_4}$ & MPM, DW/SHG at $\lambda_p/2$ & 1510 & 200 & 62 & 25 & No \\

        \cite{volpini_cladding_2026} & GaN & MPM, DW/SHG at $\lambda_p/2$ & 1560 & 120 & 22 & 30 & No\\

        \cite{brodnik_monolithic_2026} & $\mathrm{Ta_2O_5}$+TFLN & QPM, DW/SHG at $\lambda_p/2$ & 1560 & 150 & $\sim244^a$ & $\sim20$ & No\\

        \cite{chi_chip-based_nodate} & z-cut TFLN & MPM, DW/SHG at $\lambda_p/2$ & 1560 & 100 & 100 & 34 & No\\
        
        \cite{yu_coherent_2019} & $x$-cut TFLN & Not phase matched, DW/SHG at $\lambda_p/2$ & 1506 & 160 & 185 & 30 & No\\

        \cite{wu_visible--ultraviolet_2024} & $x$-cut TFLN & QPM, higher harmonics at 620 and 400 nm & 1550 & 130 & $\sim90$ & 35 & No\\
        
        \cite{hickstein_ultrabroadband_2017} & AlN & MPM, DW/SHG at $\lambda_p/2$ & 1560 & 80 & 800 & 37 & Yes\\

        \cite{okawachi_chip-based_2020} & $x$-cut TFLN & Not phase matched, DW/SHG at $\lambda_p/2$ & 1530 & 90 & 107 & 39 & Yes\\

        \textbf{\boldmath This work} & \textbf{\boldmath $x$-cut TFLN} & \textbf{\boldmath QPM, dual DWs at 1--2~\um} & \textbf{\boldmath 1579} & \textbf{\boldmath 69} & \textbf{\boldmath 16.8} & \textbf{\boldmath 32} & \textbf{\boldmath Yes}\\

        \hline

        \cite{hamrouni_picojoule-level_2024} & $x$-cut TFLN & GVM + QPM$^b$ & 2090 & 40 & 45 & 13 & No\\

        \cite{jankowski_ultrabroadband_2020} & $x$-cut TFLN & GVM + QPM$^b$ & 2050 & 50 & 11 & $35$ & No \\

        \cite{jankowski_supercontinuum_2023} & $x$-cut TFLN & GVM + QPM$^b$ & 2090 & 50 & 9.75 & $\sim35$ & No \\

        \hline\hline
    \end{tabularx}
    \caption{
    Comparison with the previous fully monolithic waveguides for $f_\text{ceo}$ detection. MPM: modal phase-matching, QPM: quasi-phase-matching, GVM: group-velocity-matched.
    \newline $^a$: Extracted from the parameters provided. 
    \newline $^b$: Utilized $\chi^{(2)}$-induced spectral broadening (cascaded/saturated second-order nonlinear interactions). 
    \newline $^c$: A concurrent work demonstrated $f_\text{ceo}$ beatnote detection on a $\mathrm{Si_3N_4}$ waveguide with a heterogeneously integrated TFLN layer (non-phase-matched SHG), achieving 30 dB SNR at 300 kHz RBW with 75 pJ pulse energy and 100 fs pulses at 1560 nm \cite{fan_f-2f_2026}.
    }
    \label{tab:device comparison v2}
\end{table}

\begin{backmatter}
\bmsection{Funding}
Defense Advanced Research Projects Agency (HR001123S0052, D23AP00252-02); Bakar Fellows Program.

\bmsection{Acknowledgment}
This work is supported by the DARPA INSPIRED program (HR001123S0052), the DARPA Young Faculty Award (D23AP00252-02), and the Bakar Fellows Program. Device fabrication was performed at the John O'Brien Nanofabrication Laboratory at the University of Southern California, the Marvell Nanofabrication Laboratory at the University of California, Berkeley, and the Molecular Foundry. Work at the Molecular Foundry was supported by the Office of Science, Office of Basic Energy Sciences, of the U.S. Department of Energy under Contract No. DE-AC02-05CH11231. The views, opinions and/or findings expressed are those of the authors and should not be interpreted as representing the official views or policies of the Department of Defense or the U.S. Government.

\bmsection{Disclosures}
C.-H.L., T.S.K., and M.Y. are involved in developing lithium niobate technologies at Opticore Inc.

\bmsection{Data availability}
Data underlying the results presented in this paper may be obtained from the authors upon reasonable request.
\end{backmatter}

\bibliography{references_final}

@article{wu_amplifier-free_2026,
  title = {Amplifier-free erbium-fiber frequency comb},
  volume = {51},
  copyright = {© 2026 Optica Publishing Group},
  issn = {1539-4794},
  url = {https://opg.optica.org/ol/abstract.cfm?uri=ol-51-14-3921},
  doi = {10.1364/OL.597507},
  language = {EN},
  number = {14},
  urldate = {2026-10-28},
  journal = {Optics Letters},
  publisher = {Optica Publishing Group},
  author = {Wu, Tsung-Han and Newman, Zachary L. and Mackenzie, Miles and Chang, Peter and Brooks, Nathan J. and Chang, Kristina F. and Hickstein, Daniel D. and Carlson, David R. and Diddams, Scott A.},
  month = jul,
  year = {2026},
  pages = {3921--3924},
}

@article{kippenberg_dissipative_2018,
	title = {Dissipative {Kerr} solitons in optical microresonators},
	volume = {361},
	url = {https://www.science.org/doi/10.1126/science.aan8083},
	doi = {10.1126/science.aan8083},
	number = {6402},
	journal = {Science},
	author = {Kippenberg, Tobias J. and Gaeta, Alexander L. and Lipson, Michal and Gorodetsky, Michael L.},
	year = {2018},
	pages = {eaan8083},
}

@article{guo_ultrafast_2023,
	title = {Ultrafast mode-locked laser in nanophotonic lithium niobate},
	volume = {382},
	url = {https://www.science.org/doi/full/10.1126/science.adj5438},
	doi = {10.1126/science.adj5438},
	number = {6671},
	urldate = {2026-10-16},
	journal = {Science},
	publisher = {American Association for the Advancement of Science},
	author = {Guo, Qiushi and Gutierrez, Benjamin K. and Sekine, Ryoto and Gray, Robert M. and Williams, James A. and Ledezma, Luis and Costa, Luis and Roy, Arkadev and Zhou, Selina and Liu, Mingchen and Marandi, Alireza},
	month = nov,
	year = {2023},
	pages = {708--713},
}

@misc{fan_f-2f_2026,
	title = {\$f\$-2\$f\$ self-referencing for silicon nitride photonics via a heterogeneously integrated lithium niobate layer},
	url = {http://arxiv.org/abs/2609.23328},
	doi = {10.48550/arXiv.2609.23328},
	urldate = {2026-10-12},
	publisher = {arXiv},
	author = {Fan, Weichen and Ruhnke, Bastian and Beeck, Camiel Op de and Tavani, Giulio and Vasiliev, Anton and Kärtner, Franz and Herr, Tobias},
	month = sep,
	year = {2026},
	note = {arXiv:2609.23328 [physics.optics]},
}

@article{brodnik_monolithic_2026,
	title = {Monolithic {3D} integration of tantalum pentoxide nonlinear photonics},
	volume = {652},
	copyright = {2026 This is a U.S. Government work and not under copyright protection in the US; foreign copyright protection may apply},
	issn = {1476-4687},
	url = {https://www.nature.com/articles/s41586-026-10379-w},
	doi = {10.1038/s41586-026-10379-w},
	language = {en},
	number = {8111},
	urldate = {2026-10-11},
	journal = {Nature},
	publisher = {Nature Publishing Group},
	author = {Brodnik, Grant M. and Spektor, Grisha and Williams, Lindell M. and Zang, Jizhao and Carollo, Alexa R. and Dan, Atasi and Black, Jennifer A. and Carlson, David R. and Papp, Scott B.},
	month = apr,
	year = {2026},
	pages = {899--905},
}

@article{tang_-chip_2026,
	title = {On-{Chip} {Amplification}-{Free} \textit{f}$_{\textrm{{CEO}}}$ {Detection} in {Parabolically} {Width}-{Modulated} {TFLN} {Waveguides}},
	volume = {44},
	copyright = {© 2026 IEEE},
	url = {https://opg.optica.org/jlt/abstract.cfm?uri=jlt-44-10-4113},
	language = {EN},
	number = {10},
	urldate = {2026-10-11},
	journal = {Journal of Lightwave Technology},
	publisher = {IEEE},
	author = {Tang, Tianyou and Yu, Simin and Zhou, Ruixin and Zhang, Jianing and Lu, Juanjuan and Chen, Guanyu and Zhu, Tao and Wang, Lingfang},
	month = may,
	year = {2026},
	pages = {4113--4119},
}

@article{chi_chip-based_nodate,
	title = {Chip-{Based} f-2f {Interferometry} in {Periodically} {Tapered} {Lithium} {Niobate} {Nanophotonic} {Waveguides}},
	volume = {n/a},
	copyright = {© 2026 Wiley-VCH GmbH},
	issn = {1863-8899},
	url = {https://onlinelibrary.wiley.com/doi/abs/10.1002/lpor.71651},
	doi = {10.1002/lpor.71651},
	language = {en},
	number = {n/a},
	urldate = {2026-10-11},
	journal = {Laser \& Photonics Reviews},
	author = {Chi, Xinyan and Yang, Ruoao and Li, Zhiyuan and Liu, Tuo and Zhang, Haoxuan and Zhan, Biyan and Liu, Xianwen},
	pages = {e71651},
}

@article{klenner_gigahertz_2016,
	title = {Gigahertz frequency comb offset stabilization based on supercontinuum generation in silicon nitride waveguides},
	volume = {24},
	copyright = {https://doi.org/10.1364/OA\_License\_v1\#VOR-OA},
	issn = {1094-4087},
	url = {https://opg.optica.org/abstract.cfm?URI=oe-24-10-11043},
	doi = {10.1364/OE.24.011043},
	language = {en},
	number = {10},
	urldate = {2026-10-06},
	journal = {Optics Express},
	author = {Klenner, Alexander and Mayer, Aline S. and Johnson, Adrea R. and Luke, Kevin and Lamont, Michael R. E. and Okawachi, Yoshitomo and Lipson, Michal and Gaeta, Alexander L. and Keller, Ursula},
	month = may,
	year = {2016},
	pages = {11043},
}

@article{hamrouni_picojoule-level_2024,
	title = {Picojoule-level supercontinuum generation in thin-film lithium niobate on sapphire},
	volume = {32},
	issn = {1094-4087},
	url = {https://opg.optica.org/abstract.cfm?URI=oe-32-7-12004},
	doi = {10.1364/OE.514649},
	language = {en},
	number = {7},
	urldate = {2026-10-06},
	journal = {Optics Express},
	author = {Hamrouni, Marin and Jankowski, Marc and Hwang, Alexander Y. and Flemens, Noah and Mishra, Jatadhari and Langrock, Carsten and Safavi-Naeini, Amir H. and Fejer, Martin M. and Südmeyer, Thomas},
	month = mar,
	year = {2024},
	pages = {12004},
}

@article{okawachi_chip-based_2020,
	title = {Chip-based self-referencing using integrated lithium niobate waveguides},
	volume = {7},
	issn = {2334-2536},
	url = {https://opg.optica.org/optica/abstract.cfm?URI=optica-7-6-702},
	doi = {10.1364/OPTICA.392363},
	language = {en},
	number = {6},
	journal = {Optica},
	author = {Okawachi, Yoshitomo and Yu, Mengjie and Desiatov, Boris and Kim, Bok Young and Hansson, Tobias and Lončar, Marko and Gaeta, Alexander L.},
	year = {2020},
	pages = {702--707},
}

@article{okawachi_carrier_2018,
	title = {Carrier envelope offset detection via simultaneous supercontinuum and second-harmonic generation in a silicon nitride waveguide},
	volume = {43},
	issn = {0146-9592, 1539-4794},
	url = {https://opg.optica.org/abstract.cfm?URI=ol-43-19-4627},
	doi = {10.1364/OL.43.004627},
	language = {en},
	number = {19},
	urldate = {2026-10-06},
	journal = {Optics Letters},
	author = {Okawachi, Yoshitomo and Yu, Mengjie and Cardenas, Jaime and Ji, Xingchen and Klenner, Alexander and Lipson, Michal and Gaeta, Alexander L.},
	month = oct,
	year = {2018},
	pages = {4627},
}

@article{carlson_self-referenced_2017,
	title = {Self-referenced frequency combs using high-efficiency silicon-nitride waveguides},
	volume = {42},
	issn = {0146-9592, 1539-4794},
	url = {https://opg.optica.org/abstract.cfm?URI=ol-42-12-2314},
	doi = {10.1364/OL.42.002314},
	language = {en},
	number = {12},
	urldate = {2026-10-06},
	journal = {Optics Letters},
	author = {Carlson, David R. and Hickstein, Daniel D. and Lind, Alex and Droste, Stefan and Westly, Daron and Nader, Nima and Coddington, Ian and Newbury, Nathan R. and Srinivasan, Kartik and Diddams, Scott A. and Papp, Scott B.},
	month = jun,
	year = {2017},
	pages = {2314},
}

@article{cormier_self-referencing_2005,
	title = {Self-{Referencing}, {Spectrally}, or {Spatially} {Encoded} {Spectral} {Interferometry} for the {Complete} {Characterization} of {Attosecond} {Electromagnetic} {Pulses}},
	volume = {94},
	url = {https://link.aps.org/doi/10.1103/PhysRevLett.94.033905},
	doi = {10.1103/PhysRevLett.94.033905},
	number = {3},
	urldate = {2026-09-29},
	journal = {Physical Review Letters},
	publisher = {American Physical Society},
	author = {Cormier, Eric and Walmsley, Ian A. and Kosik, Ellen M. and Wyatt, Adam S. and Corner, Laura and DiMauro, Louis F.},
	month = jan,
	year = {2005},
	pages = {033905},
}

@article{cundiff_colloquium_2003,
	title = {Colloquium: {Femtosecond} optical frequency combs},
	volume = {75},
	shorttitle = {Colloquium},
	url = {https://link.aps.org/doi/10.1103/RevModPhys.75.325},
	doi = {10.1103/RevModPhys.75.325},
	number = {1},
	urldate = {2026-09-29},
	journal = {Reviews of Modern Physics},
	publisher = {American Physical Society},
	author = {Cundiff, Steven T. and Ye, Jun},
	month = mar,
	year = {2003},
	pages = {325--342},
}

@article{diddams_evolving_2010,
	title = {The evolving optical frequency comb [{Invited}]},
	volume = {27},
	copyright = {© 2010 U.S. Government},
	issn = {1520-8540},
	url = {https://opg.optica.org/josab/abstract.cfm?uri=josab-27-11-B51},
	doi = {10.1364/JOSAB.27.000B51},
	language = {EN},
	number = {11},
	urldate = {2026-09-29},
	journal = {JOSA B},
	publisher = {Optica Publishing Group},
	author = {Diddams, Scott A.},
	month = nov,
	year = {2010},
	pages = {B51--B62},
}

@article{udem_optical_2002,
	title = {Optical frequency metrology},
	volume = {416},
	copyright = {2002 Springer Nature Limited},
	issn = {1476-4687},
	url = {https://www.nature.com/articles/416233a},
	doi = {10.1038/416233a},
	language = {en},
	number = {6877},
	urldate = {2026-09-29},
	journal = {Nature},
	publisher = {Nature Publishing Group},
	author = {Udem, Th and Holzwarth, R. and Hänsch, T. W.},
	month = mar,
	year = {2002},
	pages = {233--237},
}

@misc{ren_few-cycle_2026,
	title = {Few-cycle electro-optic light on thin-film lithium niobate},
	url = {http://arxiv.org/abs/2608.11393},
	doi = {10.48550/arXiv.2608.11393},
	urldate = {2026-09-28},
	publisher = {arXiv},
	author = {Ren, Xinyi and Lee, Chun-Ho and Christen, Ian and Cheung, Clayton and Kopparapu, Reshma and Yu, Yue and Zhou, Lian and Chen, Zaijun and Yu, Mengjie},
	month = aug,
	year = {2026},
	note = {arXiv:2608.11393 [physics.optics]},
}

@article{yu_integrated_2022,
	title = {Integrated femtosecond pulse generator on thin-film lithium niobate},
	volume = {612},
	copyright = {2022 The Author(s), under exclusive licence to Springer Nature Limited},
	issn = {1476-4687},
	url = {https://www.nature.com/articles/s41586-022-05345-1},
	doi = {10.1038/s41586-022-05345-1},
	language = {en},
	number = {7939},
	urldate = {2026-09-28},
	journal = {Nature},
	publisher = {Nature Publishing Group},
	author = {Yu, Mengjie and Barton III, David and Cheng, Rebecca and Reimer, Christian and Kharel, Prashanta and He, Lingyan and Shao, Linbo and Zhu, Di and Hu, Yaowen and Grant, Hannah R. and Johansson, Leif and Okawachi, Yoshitomo and Gaeta, Alexander L. and Zhang, Mian and Lončar, Marko},
	month = dec,
	year = {2022},
	pages = {252--258},
}

@article{volpini_cladding_2026,
	title = {Cladding based {GaN} waveguide engineering for efficient octave-spanning spectra generation and  fceo detection},
	volume = {128},
	issn = {0003-6951},
	url = {https://doi.org/10.1063/5.0325171},
	doi = {10.1063/5.0325171},
	number = {26},
	urldate = {2026-09-22},
	journal = {Applied Physics Letters},
	author = {Volpini, A. and Rousseau, I. and Lecomte, S. and Grassani, D.},
	month = jun,
	year = {2026},
	pages = {261104},
}

@article{arduin_doubling_2026,
	title = {Doubling the bandwidth of low-noise supercontinuum through fiber-integrated linear pulse compression},
	volume = {13},
	copyright = {© 2026 Optica Publishing Group},
	issn = {2334-2536},
	url = {https://opg.optica.org/optica/abstract.cfm?uri=optica-13-6-1098},
	doi = {10.1364/OPTICA.595939},
	language = {EN},
	number = {6},
	urldate = {2026-09-01},
	journal = {Optica},
	publisher = {Optica Publishing Group},
	author = {Arduin, Andrea and S, Shreesha Rao D. and Skov, Andreas Baltzer and Bang, Ole},
	month = jun,
	year = {2026},
	pages = {1098--1103},
}

@article{lee_inverse_2024,
	title = {Inverse design of coherent supercontinuum generation using free-form nanophotonic waveguides},
	volume = {9},
	issn = {2378-0967},
	url = {https://doi.org/10.1063/5.0196434},
	doi = {10.1063/5.0196434},
	number = {6},
	urldate = {2026-09-01},
	journal = {APL Photonics},
	author = {Lee, Chia-Yi and Liu, Yanwu and Cheng, Yinke and Lao, Chenghao and Yang, Qi-Fan},
	month = jun,
	year = {2024},
	pages = {066108},
}

@article{lamb_optical-frequency_2018,
	title = {Optical-{Frequency} {Measurements} with a {Kerr} {Microcomb} and {Photonic}-{Chip} {Supercontinuum}},
	volume = {9},
	url = {https://link.aps.org/doi/10.1103/PhysRevApplied.9.024030},
	doi = {10.1103/PhysRevApplied.9.024030},
	number = {2},
	urldate = {2026-09-01},
	journal = {Physical Review Applied},
	publisher = {American Physical Society},
	author = {Lamb, Erin S. and Carlson, David R. and Hickstein, Daniel D. and Stone, Jordan R. and Diddams, Scott A. and Papp, Scott B.},
	month = feb,
	year = {2018},
	pages = {024030},
}

@article{hickstein_ultrabroadband_2017,
	title = {Ultrabroadband {Supercontinuum} {Generation} and {Frequency}-{Comb} {Stabilization} {Using} {On}-{Chip} {Waveguides} with {Both} {Cubic} and {Quadratic} {Nonlinearities}},
	volume = {8},
	doi = {10.1103/PhysRevApplied.8.014025},
	number = {1},
	journal = {Physical Review Applied},
	author = {Hickstein, Daniel D.},
	year = {2017},
}

@article{gayer_temperature_2008,
	title = {Temperature and wavelength dependent refractive index equations for {MgO}-doped congruent and stoichiometric {LiNbO3}},
	volume = {91},
	issn = {1432-0649},
	url = {https://doi.org/10.1007/s00340-008-2998-2},
	doi = {10.1007/s00340-008-2998-2},
	language = {en},
	number = {2},
	urldate = {2026-07-28},
	journal = {Applied Physics B},
	author = {Gayer, O. and Sacks, Z. and Galun, E. and Arie, A.},
	month = may,
	year = {2008},
	pages = {343--348},
}

@article{jankowski_ultrabroadband_2020,
	title = {Ultrabroadband nonlinear optics in nanophotonic periodically poled lithium niobate waveguides},
	volume = {7},
	copyright = {© 2020 Optical Society of America},
	issn = {2334-2536},
	url = {https://opg.optica.org/optica/abstract.cfm?uri=optica-7-1-40},
	doi = {10.1364/OPTICA.7.000040},
	language = {EN},
	number = {1},
	urldate = {2026-07-23},
	journal = {Optica},
	publisher = {Optica Publishing Group},
	author = {Jankowski, Marc and Langrock, Carsten and Desiatov, Boris and Marandi, Alireza and Wang, Cheng and Zhang, Mian and Phillips, Christopher R. and Lončar, Marko and Fejer, M. M.},
	month = jan,
	year = {2020},
	pages = {40--46},
}

@article{carlson_generating_2019,
	title = {Generating few-cycle pulses with integrated nonlinear photonics},
	volume = {27},
	issn = {1094-4087},
	url = {https://opg.optica.org/abstract.cfm?URI=oe-27-26-37374},
	doi = {10.1364/OE.27.037374},
	language = {en},
	number = {26},
	urldate = {2026-06-22},
	journal = {Optics Express},
	author = {Carlson, David R. and Hutchison, Phillips and Hickstein, Daniel D. and Papp, Scott B.},
	month = dec,
	year = {2019},
	pages = {37374},
}

@article{wu_visible--ultraviolet_2024,
	title = {Visible-to-ultraviolet frequency comb generation in lithium niobate nanophotonic waveguides},
	volume = {18},
	copyright = {2024 The Author(s), under exclusive licence to Springer Nature Limited},
	issn = {1749-4893},
	url = {https://www.nature.com/articles/s41566-023-01364-0},
	doi = {10.1038/s41566-023-01364-0},
	language = {en},
	number = {3},
	urldate = {2026-06-11},
	journal = {Nature Photonics},
	publisher = {Nature Publishing Group},
	author = {Wu, Tsung-Han and Ledezma, Luis and Fredrick, Connor and Sekhar, Pooja and Sekine, Ryoto and Guo, Qiushi and Briggs, Ryan M. and Marandi, Alireza and Diddams, Scott A.},
	month = mar,
	year = {2024},
	pages = {218--223},
}

@article{yu_coherent_2019,
	title = {Coherent two-octave-spanning supercontinuum generation in lithium-niobate waveguides},
	volume = {44},
	issn = {0146-9592, 1539-4794},
	url = {https://opg.optica.org/abstract.cfm?URI=ol-44-5-1222},
	doi = {10.1364/OL.44.001222},
	language = {en},
	number = {5},
	urldate = {2026-05-14},
	journal = {Optics Letters},
	author = {Yu, Mengjie and Desiatov, Boris and Okawachi, Yoshitomo and Gaeta, Alexander L. and Lončar, Marko},
	month = mar,
	year = {2019},
	pages = {1222},
}

@article{jankowski_supercontinuum_2023,
	title = {Supercontinuum generation by saturated second-order nonlinear interactions},
	volume = {8},
	issn = {2378-0967},
	url = {https://doi.org/10.1063/5.0158926},
	doi = {10.1063/5.0158926},
	number = {11},
	urldate = {2026-05-14},
	journal = {APL Photonics},
	author = {Jankowski, Marc and Langrock, Carsten and Desiatov, Boris and Lončar, Marko and Fejer, M. M.},
	month = nov,
	year = {2023},
	pages = {116104},
}

@article{genty_nonlinear_2007,
	title = {Nonlinear envelope equation modeling of sub-cycle dynamics and harmonic generation in nonlinear waveguides},
	volume = {15},
	copyright = {© 2007 Optical Society of America},
	issn = {1094-4087},
	url = {https://opg.optica.org/oe/abstract.cfm?uri=oe-15-9-5382},
	doi = {10.1364/OE.15.005382},
	language = {EN},
	number = {9},
	urldate = {2026-05-12},
	journal = {Optics Express},
	publisher = {Optica Publishing Group},
	author = {Genty, G. and Kinsler, P. and Kibler, B. and Dudley, J. M.},
	month = apr,
	year = {2007},
	pages = {5382--5387},
}

@article{wabnitz_harmonic_2010,
	title = {Harmonic and supercontinuum generation in quadratic and cubic nonlinear optical media},
	volume = {27},
	copyright = {© 2010 Optical Society of America},
	issn = {1520-8540},
	url = {https://opg.optica.org/josab/abstract.cfm?uri=josab-27-9-1707},
	doi = {10.1364/JOSAB.27.001707},
	language = {EN},
	number = {9},
	urldate = {2026-05-11},
	journal = {JOSA B},
	publisher = {Optica Publishing Group},
	author = {Wabnitz, S. and Kozlov, V. V.},
	month = sep,
	year = {2010},
	pages = {1707--1711},
}

@article{manurkar_fully_2018,
	title = {Fully self-referenced frequency comb consuming 5 watts of electrical power},
	volume = {1},
	issn = {2578-7519},
	url = {https://opg.optica.org/osac/abstract.cfm?uri=osac-1-1-274},
	doi = {10.1364/OSAC.1.000274},
	language = {EN},
	number = {1},
	urldate = {2026-10-28},
	journal = {OSA Continuum},
	publisher = {Optica Publishing Group},
	author = {Manurkar, Paritosh and Perez, Edgar F. and Hickstein, Daniel D. and Carlson, David R. and Chiles, Jeff and Westly, Daron A. and Baumann, Esther and Diddams, Scott A. and Newbury, Nathan R. and Srinivasan, Kartik and Papp, Scott B. and Coddington, Ian},
	month = sep,
	year = {2018},
	pages = {274--282},
}

\end{document}